\documentclass[aps,pre,twocolumn,preprintnumbers,amsmath,amssymb,a4paper,10pt,nofootinbib]{revtex4-1}
\usepackage{graphicx}
\usepackage[caption=false,subrefformat=parens,labelformat=parens]{subfig}
\usepackage{url}
\usepackage{rotating}
\usepackage{xcolor}
\usepackage{txfonts}

\def\LC{$L$-cloned~}
\def\X{$L_\mathbb{X}$}
\newcommand{\ed}[1]{{\color{black} #1}}

\usepackage[top=2.5cm,bottom=2.2cm,left=1.6cm,right=1.6cm]{geometry}

\makeatletter
\g@addto@macro\bfseries{\boldmath}
\makeatother

\begin{document}
\title{Network cloning unfolds the effect of clustering on dynamical processes}
\author{Ali Faqeeh, Sergey Melnik, James P. Gleeson}
\affiliation{MACSI, Department of Mathematics \& Statistics, University of Limerick, Ireland}

\begin{abstract}
We introduce network $L$-cloning, a technique for creating ensembles of random networks from any given real-world or artificial network. Each member of the ensemble is an $L$-cloned network constructed from $L$ copies of the original network. The degree distribution of an $L$-cloned network and, more importantly, the degree-degree correlation between and beyond nearest neighbors are identical to those of the original network. The density of triangles in an \LC network, and hence its clustering coefficient, is reduced by a factor of $L$ compared to those of the original network. Furthermore, the density of loops of any fixed length approaches zero for sufficiently large values of $L$. Other variants of $L$-cloning allow us to keep intact the short loops of certain lengths. As an application, we employ these network cloning methods to investigate the effect of short loops on dynamical processes running on networks and to inspect the accuracy of corresponding tree-based theories. We demonstrate that dynamics on $L$-cloned networks (with sufficiently large $L$) are accurately described by the so-called adjacency tree-based theories, examples of which include the message passing technique, some pair approximation methods, and the belief propagation algorithm  used respectively to study bond percolation, SI epidemics, and the Ising model.
\end{abstract}
\maketitle

\section{Introduction} \label{sec1}
The behavior of processes such as percolation (as a model for network resilience) or susceptible-infected (SI) disease spreading depends on the structure of the underlying network on which they operate \cite{Newman10}. Degree distribution and degree-degree correlation are two important structural properties of a network that influence its dynamics \cite{Newman10}. Moreover, the presence of an appreciable number of short loops in the network (referred to as clustering) is known to significantly affect dynamics \cite{Newman10,Newman09,Miller09b,Gleeson09a}. Real-world networks often have a relatively large clustering coefficient \cite{Newman10,Boccaletti06,Newman03a}, hence they possess a relatively large number of loops of length 3 (i.e., simple cycles of length 3). Some recent network models are able to produce random networks with a desired number of short loops \cite{Gleeson09a,Karrer82,Gleeson09b,Hackett12} and provide theoretical frameworks for the analysis of bond and site percolation properties, as well as calculating the sizes of $k$-cores and global cascades. However, the state-of-the-art theoretical methods cannot capture the effect of clustering on dynamics running on real-world networks.

Commonly used theories for dynamical processes running on real-world networks are tree-based, i.e., they assume (in one way or another) that the network has a locally treelike structure. Examples of tree-based theories include mean-field theories (which represent the network by its degree distribution or degree-degree correlation) and a class of pair-approximation methods \cite{Melnik11,Gleeson12,Gleeson13}. More sophisticated and accurate tree-based theories (which we call adjacency tree-based or $A_{ij}$ theories for short) use information on the adjacency of individual nodes. These theories have been used to study different dynamical processes including site percolation on directed networks \cite{restrepo2008weighted}, bond percolation \cite{karrer_PRL14} and SI disease spread (see Sec.~17.10 of Ref.~\cite{Newman10}). The belief propagation, the Bethe-Peierls, and the cavity methods (used, for example, to study the Ising model)~\cite{Dorogovtsev08,Shrestha_PRE14} are also $A_{ij}$ theories. Although tree-based methods are expected to fail on clustered networks, they perform reasonably well on some clustered networks, which casts doubts on the origins of inaccuracy~\cite{Melnik11}.
\ed{We have previously demonstrated that tree-based theories perform worse on clustered networks that have low values of average degree and nearest neighbors degree~\cite{Gleeson12}. This study showed that there are important characteristics other than the network clustering coefficient that affect the accuracy of the tree-based theories. However, the net effect of the short loops on the accuracy of theories was not addressed.
}

In this paper we investigate how short loops influence the network dynamics and inspect the accuracy of existing theories for dynamical processes operating on real-world networks. In this regard, we propose the \LC networks that can be constructed for any given real-world or artificial network via a simple process that we call $L$-cloning. We show that the ensemble of $L$-cloned networks mimics some of the important properties of the original network, including degree distribution and degree-degree correlation between and beyond nearest neighbors, while the network clustering coefficient is reduced by a factor of $L$ compared to that of the original network.
\ed{We also introduce an extension of the $L$-cloning method in which densities of loops of specified lengths are also preserved. We use this extension to highlight the net effect of short loops on network dynamics.} These features make network cloning an ideal framework for isolating the effect of short loops on dynamical processes and for evaluating the performance of theoretical models in the presence of network clustering.

This paper is organized as follows. In Sec.~\ref{sec2}, we introduce and discuss the design and some of the structural properties of networks constructed by $L$-cloning and its extensions. In Sec.~\ref{sec3}, we consider several dynamical processes operating on networks and their corresponding $A_{ij}$ theories. We then investigate the effect of short loops on dynamical processes and the accuracy of corresponding theoretical predictions by applying network cloning methods to some synthetic and real-world clustered networks. In Sec.~\ref{sec4}, we present a summary and point out potential future work and possible applications of $L$-cloning.

\section{Cloned networks} \label{sec2}
In this section, we first describe the ensemble of \LC networks and explain how they can be constructed from a network of interest (e.g., a real-world network). We then describe their structural properties, focusing on how short loops in the network are affected by the $L$-cloning. \ed{Afterward, we introduce two extensions of $L$-cloning, i.e., \X-cloning, and $L_f$-cloning (defined in Secs.~\ref{sec2d} and \ref{sec2e}).}

The main goal of proposing the $L$-cloned networks ensemble is to have networks that are very similar, in most respects, to the network of interest, while having the effect of short loops reduced or practically removed. As we show, this isolates the effect of clustering from other structural properties of the network, and assists in understanding how networked behavior can be influenced solely by clustering.

In the remainder of this section we give a detailed description of the construction and properties of \LC networks, but to summarize briefly, \LC networks have the following features:
\begin{enumerate}
  \item An \LC network can be constructed for any given network and any positive integer $L$.
  \item An \LC network has $L$ times as many nodes and edges as the original network.
  \item The degree distribution and degree-degree correlation between and beyond the nearest neighbors in an \LC network are identical to those of the original network.
  \item The expected clustering coefficient of an \LC network is $L$ times smaller than that of the original network. More generally, the density of loops of any fixed length approaches zero for sufficiently large values of $L$.
  \item \ed{The first three statements above also hold for \X~and $L_f$ cloned networks. On the other hand \X-cloning preserves short loops of specified lengths.}

\end{enumerate}

\subsection{Design and description}
\begin{figure}!t]
{\includegraphics[width=1\columnwidth]{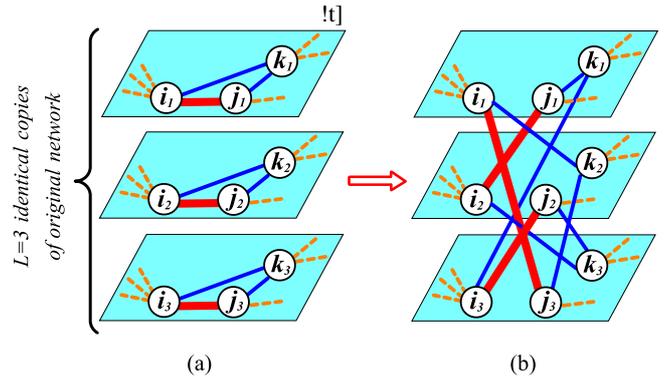}}
\caption{(Color online) (a) To create an \LC network, we start with $L$ clones of the original network. In particular, for each node $i$ of the original network, there exist $L$ copies $i_1, i_2, ..., i_L$ each placed in one of the $L$ layers. (b) We then reassign existing links uniformly at random, subject to the following restriction: If nodes $i$ and $j$ are connected in the original network, then each copy of $i$ is connected to exactly one copy of $j$ and each copy of $j$ is connected to exactly one copy of $i$.}\label{f1}
\end{figure}

In order to build an \LC network, we start with $L$ clones (i.e., identical copies) of the original network. In particular, for each node $i$ of the original network, there exist $L$ copies $i_1, i_2, ..., i_L$ each placed in one of the $L$ layers as we show in Fig.~\ref{f1}(a). An \LC network is then created by interweaving the $L$ layers by reassigning the existing links uniformly at random, subject to the following restriction: If nodes $i$ and $j$ are connected in the original network, then each copy of $i$ is connected to exactly one copy of $j$ and each copy of $j$ is connected to exactly one copy of $i$ (see Fig.~\ref{f1}(b)). The ensemble of \LC networks is comprised of realizations of this $L$-cloning procedure for a fixed number of layers $L$.

An \LC network is $L$ times larger than the original network; nevertheless it has the same degree distribution and nearest neighbors degree-degree correlation. Moreover degree-degree correlation beyond nearest neighbors is also preserved by $L$-cloning. That is, for any node in the original network, the set of degrees of its neighbors, its second neighbors, or any group of nodes at a distance $d$ from the node, is identical to that of any of its clones in the \LC network. In other words, the set of degrees of nodes that are $d$ links away from a node $i$, denoted by $\{K\}_{(i,d)}$, is identical for all clones of the original node $i$ for all $d$ values. Moreover, for each path in the original network, there exist exactly $L$ related paths in the $L$-cloned network; the sequence of nodes in each of these $L$ paths consists of identically ordered copies of nodes of the original path. Hence, such paths have identical sequences of node degrees.

By contrast, an \LC network has a different density of loops (clustering) than the original network. The density of triangles is decreased: The shuffling of links between layers in the $L$-cloned network breaks the triangles in each layer, constructing a new structure in which triangles are less likely to happen. For example, instead of $3$ triangles in Fig.~\ref{f1}(a), there is $1$ triangle ($i_1-j_3-k_2-i_1$) in Fig.~\ref{f1}(b). In fact, a fraction of the broken triangles create new triangles, while the rest of them merge into longer loops. In Fig.~\ref{f1}(b), for example, a loop of length 6 ($j_1-k_1-i_3-j_2-k_3-i_2-j_1$) is created in addition to the aforementioned new triangle. In Sec.~\ref{sec2c}, we show that the density of triangles is decreased by a factor of $L$ and discuss the effect of $L$-cloning on longer loops.

\subsection{The effect of \textit{L}-cloning on clustering} \label{sec2c}
In this section, we discuss how the density of short loops is affected by $L$-cloning. In particular, we provide an analytical expression for the change in the clustering coefficient~\cite{Boccaletti06} and verify it by numerical simulations.

Consider a triangle in the original network. Figure~\ref{f1}(a) shows $L$ copies of the triangle each located in one of the $L$ separate layers. Disregarding the node sub-indices for the moment, each of these $L$ triangles consists of a dyad (a connected subgraph consisting of 2 links and 3 nodes) of edges $i-k-j$ completed by a third (thick red) edge $i-j$. Importantly, in the \LC network depicted in Fig.~\ref{f1}(b), the number of dyads $i-k-j$ remains the same. However, the probability that each dyad is now completed by a third edge $i-j$ (and hence forms a triangle) is $1/L$ because links are reassigned at random and each copy of node $i$ can be connected to any of $L$ copies of node $j$. This means that the density of triangles in an \LC network is reduced by a factor of $L$. (Note that in an \LC network, the expected number of triangles is the same as in the original one, but an \LC network has $L$ times as many nodes and edges.)

\begin{figure}[t]
\includegraphics[width=1\columnwidth]{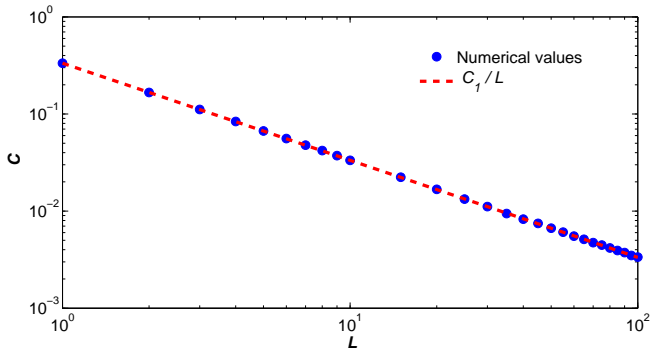}
\caption{(Color online) The dependence of clustering coefficient $C$ of $L$-cloned networks on $L$ (blue circles). Here $C_1$ is the value of the clustering coefficient of the original ($L=1$) network. The red dashed line is $C_1/L$.}\label{f2}
\end{figure}

Therefore, if the clustering coefficient of the original network is $C$, the expected clustering coefficient of its \LC version is $C/L$. This relation holds for both common definitions of the clustering coefficient, given respectively by Eqs.~(2.8) and (2.9) of Ref.~\cite{Boccaletti06}. Figure \ref{f2} illustrates the changes of the clustering coefficient (defined by Eq.~(2.9) of Ref.~\cite{Boccaletti06}) with respect to $L$ for $L$-cloned networks constructed from a $\gamma(3,3)$ network.

Similarly, the density of loops of length 4 or 5 (or any prime number above 5) in an \LC network is $L$ times smaller than that in the original network. However, the change in the density of loops of other lengths (e.g., $6$ or $8$) is more complicated and depends on both $L$ and the density of some shorter loops. The reason for this is that $L$-cloning can transform a loop of length $m$ that exists in the original networks into loops of length $m$, $2m$, ..., $mL$ in the \LC network.\footnote {To verify this, suppose that we have a loop of length 3 (a triangle) $i-j-k-i$ in the original network as shown in Fig.~\ref{f1}(a). Every loop that can be created by $L$-cloning from this one will have the same sequence of letters $i-j-k-i$, possibly repeated up to $L$ times with different subscripts (denoting the layer of the copy): $i-j-k-i-j-k-i-...-i$. Observe, for example, the loops $i_1-j_3-k_2-i_1$ and $i_2-j_1-k_1-i_3-j_2-k_3-i_2$ in Fig.~\ref{f1}(b). Conversion of triangles to loops of lengths that are not multiples of 3 would require a different sequence of connected copies of nodes, which is impossible by the definition of the $L$-cloning procedure. In general, from loops of length $m$ in the original network only loops of lengths that are multiples of $m$ can be created in the $L$-cloned ensemble.}
Nevertheless, for sufficiently large $L$ the density of loops of any fixed length will approach zero. Accordingly, by increasing $L$ one can investigate how diminishing the density of short loops in $L$-cloned networks affects networked behavior such as dynamical processes.

It is also worth mentioning that \LC networks can be regarded as multilayer networks, which were reviewed and categorized in Ref.~\cite{kivela2014JCN}. We discuss this in more detail in the Appendix. Using the multilayer representation of \LC networks is not the main focus of this paper, however it can benefit further studies.

\subsection{Extension of \textit{L}-cloning that preserves loops of certain lengths} \label{sec2d}
It is possible to modify the edge reassignment rules of the $L$-cloning process to preserve additional network characteristics. A useful and informative extension of $L$-cloning is one that preserves loops of specified lengths, e.g., triangles and/or loops of length 4 (squares). We refer to this extension of cloning as \X-cloning, where the set $\mathbb{X}$ specifies the lengths of the loops whose edges are excluded from the reassignment stage of $L$-cloning. For example, to preserve the density of triangles we exclude edges that are part of one or more triangles from the edge reassignment. Hence, only edges which are not part of any triangle are reassigned (rewired) and interweave the $L$ layers of the network. This type of cloning is accordingly referred to as $L_{-3}$-cloning. By only rewiring the edges that are not part of any triangle, $L_{-3}$-cloning keeps the triangles intact and can only change the densities of longer loops.

Similarly, to keep the density of squares it is sufficient to  avoid rewiring the edges that are part of such loops. Moreover, in order to preserve the density of both triangles and squares, the edges that are part of either of such loops are not rewired; we call this $L_{-3,-4}$-cloning. In a similar way, the cloning process can be modified to preserve the densities of loops of any chosen lengths.

It is worth mentioning that, similar to $L$-cloning, these extensions preserve the degree distribution and degree-degree correlation between and beyond nearest neighbors of the original network from which they are constructed. Additionally, some other characteristics are also preserved. For example, denote the number of triangles and squares that contain node $i$ by $t_i$ and $s_i$ respectively and consider $L_{-3,-4}$-cloning. All clones of node $i$ will have the same $t_i$ and $s_i$ values. Furthermore, the sequence of these values on any path in $L_{-3,-4}$-cloned networks is identical to that of the corresponding path in the original network.

\subsection{Partial $L$-cloning} \label{sec2e}
In this section, we introduce another extension of $L$-cloning that will be used to inspect the effect of short loops. Let $f$ denote the fraction of links that are reassigned (rewired) during network cloning. Accordingly, $f=1$ in the case of regular $L$-cloning where all the links are rewired, but $f<1$ when we $L_{-3}$-clone a network with non-zero clustering coefficient because we do not rewire links that form triangles. In order to check whether the fraction of rewired links itself has an effect on the dynamics, we propose another extension of $L$-cloning that we call $L_{f}$-cloning. For $L_f$-cloning, a fixed fraction $f$ of the links of the original network are randomly selected. Then $L_f$-cloning proceeds as the regular $L$-cloning with the difference that only the randomly selected links are reassigned: For each randomly selected link $i-j$ of the original network, the links between clones of $i$ and $j$ (located in the $L$ identical layers) are reassigned such that each clone of $i$ is connected to exactly one randomly chosen clone of $j$ and vice versa.

An $L_{f}$-cloning in which the fraction of randomly selected links to be reassigned is identical to the fraction of links reassigned in the $L_{{-3}}$-cloning (or similarly $L_{-3,-4}$-cloning) is dubbed $L_{f(3)}$-cloning (or $L_{f(3,4)}$-cloning). In $L_f$-cloning, the reassigned links are randomly selected from the set of all links,
hence they may be part of triangles or squares. Therefore, in contrast to $L_{-3}$ and $L_{-3,-4}$-cloning, $L_f$-cloning reduces the density of triangles and squares. Comparing the simulations on the different types of cloned networks (e.g., $L$-, $L_{-3}$-, $L_{-3,-4}$- and $L_f$-cloned networks) constructed from the same original network can provide insight into the effects of short loops on the network dynamics and structure.

\subsection{Clustered networks for testing the effect of \textit{L}-cloning}\label{sec2b}
\begin{figure}[ht!]
\includegraphics[width=.7\columnwidth]{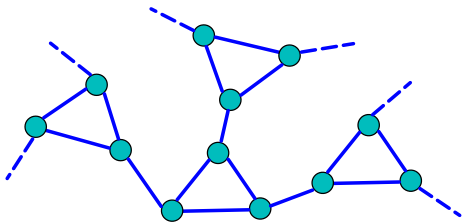}
\caption{(Color online) Schematic of a $\gamma(3,3)$ network described in the text. All nodes have degree 3. Each node is part of exactly one triangle and has exactly one single edge that is not part of any triangle. The triangles are randomly connected via single edges. The dashed lines represent connections to nodes not included in the schematic.} \label{fg33}
\end{figure}

In this section, we describe two networks, one synthetic and one empirical, that we will use in the rest of the paper to help demonstrate various points. We consider a class of synthetic clustered networks defined by the joint probability distribution $\gamma(k,c)$, which gives the probability that a randomly chosen node has degree $k$ and is a member of a $c$-clique (an all-to-all connected subgraph of $c$ nodes) \cite{Gleeson09a}. We focus on a specific case with $\gamma(3,3)=1$ (and $\gamma(k,c)=0$ for other values of $k$ and $c$), and we dub such networks $\gamma(3,3)$ networks. In $\gamma(3,3)$ networks all nodes have degree 3. Each node is part of exactly one triangle and has exactly one single edge that is not part of any triangle. The single edges randomly connect pairs of nodes from different triangles. Figure~\ref{fg33} illustrates a part of a $\gamma(3,3)$ network. We note that $\gamma(3,3)$ networks are equivalent to the $p_{1,1} = 1$ case in the clustered random graph model of Refs.~\cite{Newman09,Miller09b}, where $p_{s,t}$ is the probability that a randomly chosen node is part of $t$ different triangles and in addition has $s$ single edges (which do not belong to the triangles).

Previous studies on the accuracy of tree-based theories for dynamics running on clustered networks~\cite{Melnik11,Gleeson12} showed that mean-field and pair-approximation methods are generally inaccurate for $\gamma(3,3)$ networks. In these studies, the power grid network of the western United States~\cite{Watts98,Melnik11} was also shown to be a real-world example for which the tree-based theories have a poor performance. Accordingly, we use the $\gamma(3,3)$ and the power grid network to exemplify the effect of $L$-cloning on dynamical processes operating on clustered networks and to inspect the accuracy of corresponding $A_{ij}$ theories.

\section{Dynamical processes on networks and $A_{ij}$ theories} \label{sec3}
In this section, we consider several dynamical processes operating on networks, and inspect their behavior on cloned versions of several synthetic and real-world clustered networks. By comparing the results against theoretical predictions we investigate the effect of clustering on the accuracy of tree-based theories.

As we mentioned in Sec.~\ref{sec1}, a class of tree-based theories that use information on the adjacency of individual nodes is here referred to as $A_{ij}$ theories. In this class, the state of each node is considered to be independently related to the states of its immediate neighbors (neglecting possible effects of interaction between the neighbors). This class of theories, which can often be recognized by the appearance of the adjacency matrix (or its elements $A_{ij}$) in the governing equations, has found applications in many areas, including bond percolation \cite{Karrer10,karrer_PRL14}, SI epidemics (Sec.~17.10 of Ref.~\cite{Newman10}), and belief propagation and Bethe-Peierls methods for the Ising model~\cite{Dorogovtsev08}. The $A_{ij}$ theories are generally more accurate than their reductions to, for example, \emph{degree-based} approximations \cite{Newman10,porter2014dynamical}, which use excess degree or conditional probability distributions as approximations to the exact connection information provided by the network adjacency matrix. However, $A_{ij}$ theories are not exact on networks with short loops.

In this section, we study several dynamical processes running on networks for which $A_{ij}$ theories are used. The processes considered here are binary-state dynamics \cite{Gleeson13}, in which each node can be in one of two possible states. For these processes we demonstrate that
$A_{ij}$ theories are accurate on \LC versions of a clustered network for sufficiently large $L$. To investigate the effect of short loops on the accuracy of $A_{ij}$ theories we also use the other variants of cloning described in Secs.~\ref{sec2d} and~\ref{sec2e}.

\subsection{Percolation} \label{sec3a}
Bond and site percolation are among the most widely studied models on complex networks \cite{Newman10,Dorogovtsev08}.
In bond (site) percolation a fraction $1-p$ of the links (nodes) in the network are removed and the remaining fraction $p$ of links (nodes) constitute the structure of a new damaged (and possibly disconnected) network; $p$ is called the occupation probability. If $p$ is sufficiently large, a giant connected component (GCC) with a size that scales linearly with the network size exists; otherwise the network is collapsed into isolated small components with vanishing fractional sizes in the limit of infinite network size. The quantity of interest here is the probability $S$ that a random node is part of the GCC (i.e., the fraction of nodes in the GCC), and how it depends on the occupation probability $p$.

Karrer \emph{et al.}~\cite{karrer_PRL14} recently showed that bond percolation can be formulated as a message passing process on locally treelike networks. We employ their results to calculate the size of the GCC for bond and site percolation. Consider a network with adjacency matrix ${A}$ and define $u_{ij}$ as the probability that node $i$ is not connected to GCC via node $j$. If we then assume that the network is locally treelike, for bond (site) percolation we have
\begin{equation}\label{eq1}
  u_{ij}=1-p+p\prod_{k\neq i}{A}_{jk}u_{jk} ~,
\end{equation}
where the term $1-p$ denotes the probability that the link $i-j$ (node $j$) is not occupied. The last term in Eq.~(1) is the product of $p$, the probability that the link $i-j$ (site $j$) is occupied, and the probability that none of the other links leaving $j$ lead to the GCC. Then, for the case of bond percolation, node $i$ is not connected to GCC if none of its links lead to the GCC, which happens with probability $U_i=\prod_j {A}_{ij}u_{ij}$, hence it is in the GCC with probability $1-U_i$. For site percolation, the probability that $i$ is part of the GCC is $p(1-U_i)$ where $p$ denotes the probability that $i$ is itself occupied. Averaging over all nodes, we obtain the fractional size of the GCC for bond and site percolation, respectively,
\begin{eqnarray}
  ~~~~S=1-\frac{1}{N}\sum_i\prod_j{A}_{ij}u_{ij} ~,~~~~~~~~~~\label{eq2a}\\
  ~~~S=p\left(1-\frac{1}{N}\sum_i\prod_j{A}_{ij}u_{ij}\right) .~~~~~~~~~\label{eq2b}
\end{eqnarray}

Equation~\eqref{eq2a} resembles Eq.~(24) of Ref.~\cite{Karrer10}, which is derived for the late time behavior of the susceptible-infected recovered model using the message passing approach and is the same as Eq.~(7) of Ref.~\cite{karrer_PRL14}. This theory is an example of an $A_{ij}$ theory which uses the information on the exact connectivity of pairs of nodes represented in the network adjacency matrix. If we assume that the edges leaving nodes with degree $k$ have approximately the same probability $u_k$ of not leading to the GCC, then Eqs. (\ref{eq1}) and (\ref{eq2a}) can be reduced to a degree-based approximation as described by Eqs. (1) and (2) of Ref.~\cite{Vazquez03}:
\begin{eqnarray}
  u_k &=& 1-p+p\sum_{k'}P(k'|k)(u_{k'})^{k'-1} ~,\label{eqPkka} \\
  S &=& 1-\sum_k P(k)(u_k)^k ~,\label{eqPkkb}
\end{eqnarray}
where $P(k)$ is the degree distribution, and $P(k'|k)$ is the conditional probability that a neighbor of a degree-$k$ node is a degree-$k'$ node. This assumption is correct when the degree of neighbors of a node with any degree can be well approximated using the conditional probability; this is the case, for example, for configuration model and $P_{k,k'}$-rewired networks \cite{Melnik11}. In the absence of degree-degree correlation, all edges have the same probability of leading to the GCC and Eq.~(\ref{eq1}) is further reduced to Eq. (16.4) or (17.27) of Ref.~\cite{Newman10} and Eqs.~(\ref{eq2a}) and (\ref{eq2b}) are reduced to Eqs. (16.2) and (17.28) of \cite{Newman10} respectively.

\begin{figure}[b!] \centering
\subfloat{\includegraphics[width=1\columnwidth]{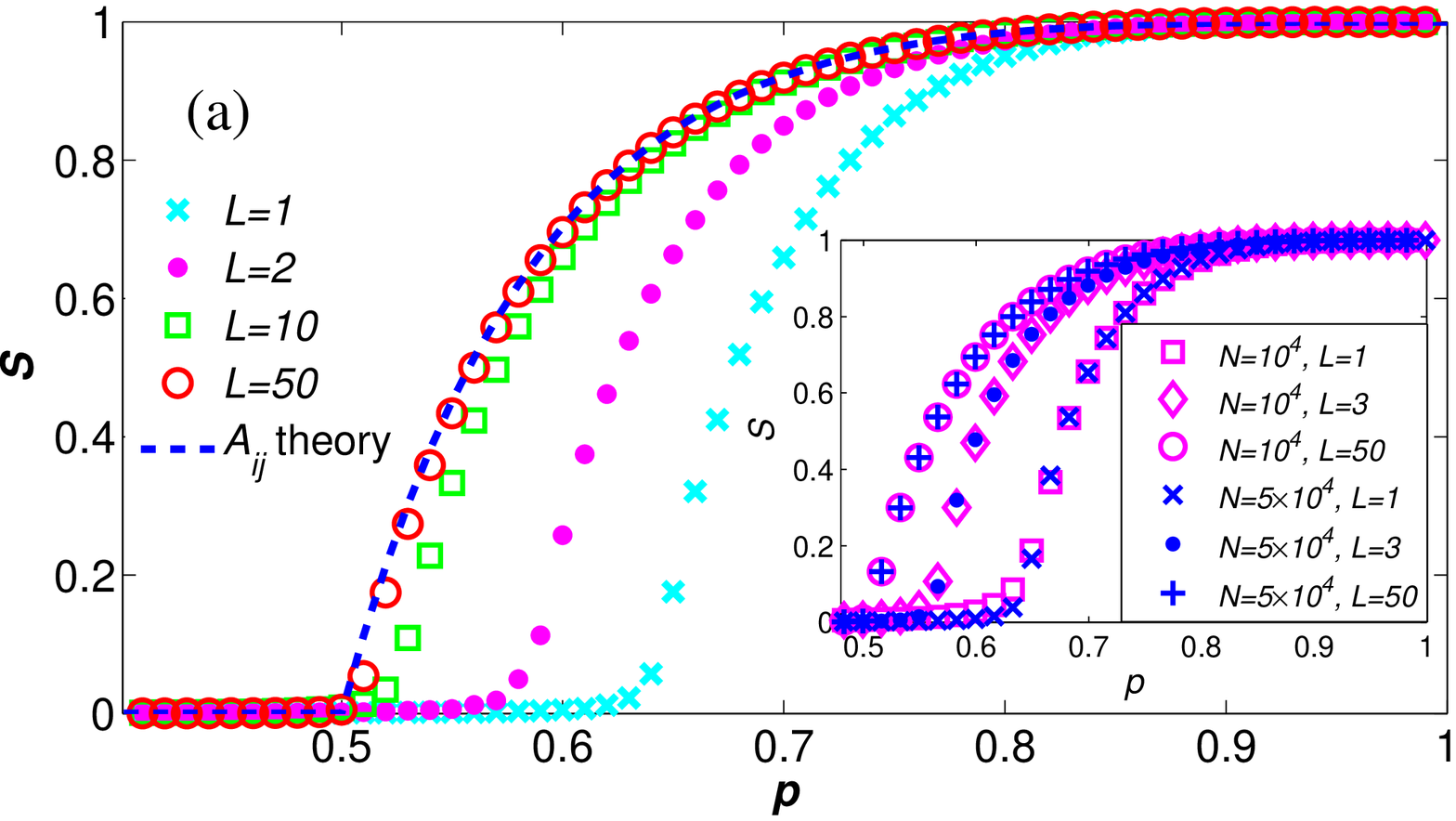}\label{fBa}}
\hfill
\subfloat{\includegraphics[width=1\columnwidth]{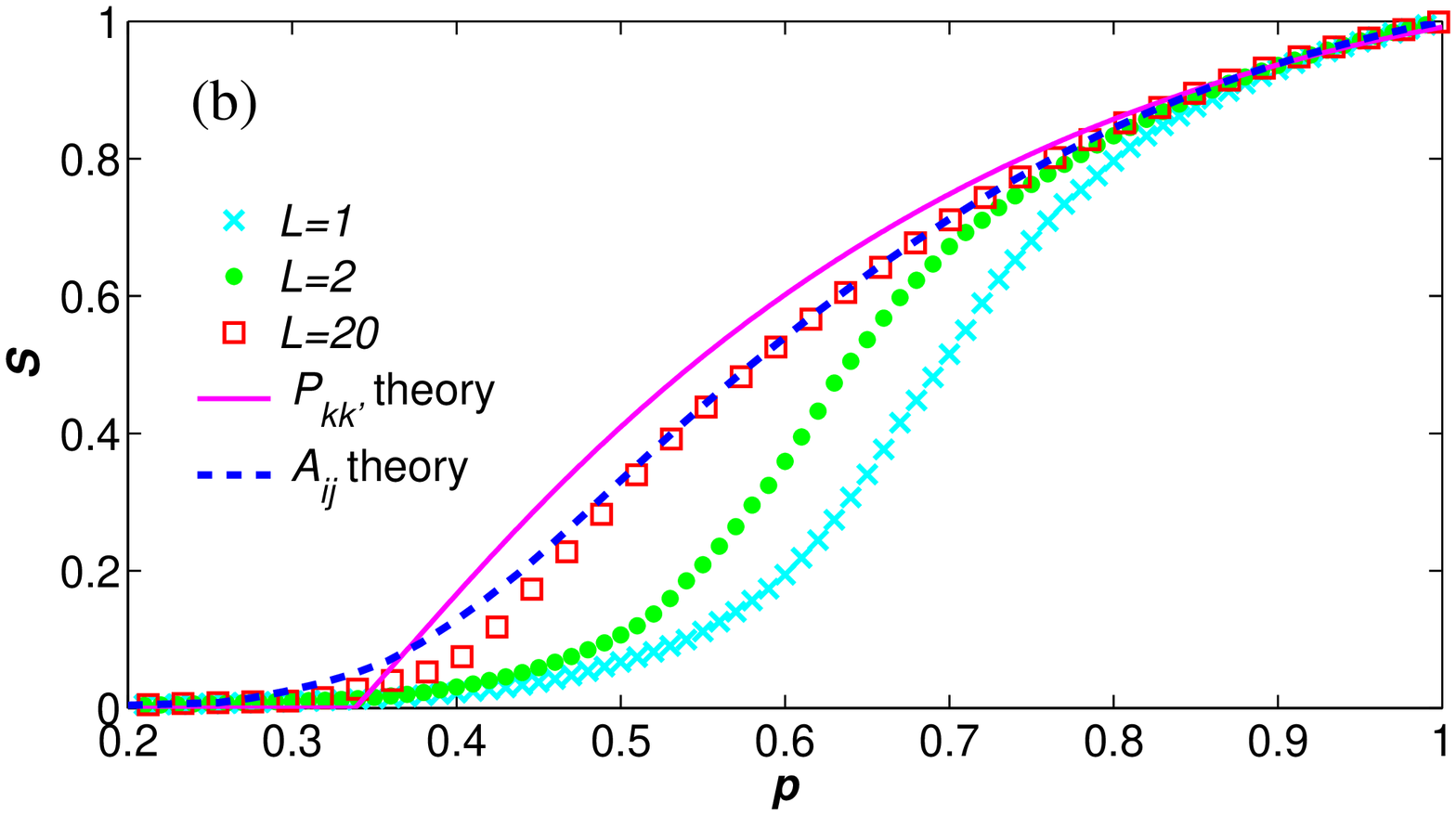}\label{fBb}}
\caption{(Color online) Bond percolation on (a) a $\gamma(3,3)$ network with approximately $10^4$ nodes, and (b) the US power grid network and their respective \LC networks. Here $L=1$ indicates the result of numerical simulations on the original network. The result of the $A_{ij}$ theory is shown by the dashed line. The result of numerical simulation on \LC networks approaches the $A_{ij}$ theory as $L$ is increased. On the US power grid network the prediction made by the $A_{ij}$ theory is different from and more accurate than that of the $P_{k,k'}$ theory studied in Refs. \cite{Vazquez03,Melnik11}. Nevertheless, the two theories have the same result on $\gamma(3,3)$ networks. The numerical results were derived from averaging over at least 100 realizations of the bond percolation process.}
\end{figure}
Although Eqs.~(\ref{eq1})$-$(\ref{eq2b}) are more general and precise than their aforementioned reductions, they still employ the assumption that the network is locally treelike. Hence, they can be inaccurate for clustered networks~\cite{Gleeson12, Melnik11}.

To inspect the role of short loops in the accuracy of the aforementioned $A_{ij}$ theory we consider synthetic and real-world clustered networks introduced in Sec.~\ref{sec2b} and apply $L$-cloning to reduce clustering while preserving degree distribution and degree-degree correlation between and beyond nearest neighbors. In Fig.~\subref*{fBa}, we show how the size of the GCC, in bond percolation, depends on $p$ for the $\gamma(3,3)$ network and its $L$-cloned versions with different $L$, and compare the numerical results with the theoretical predictions of the $A_{ij}$ theory from Eqs.~(\ref{eq1}) and (\ref{eq2a}). The theory does not make an accurate prediction for the original network ($L=1$). However, when $L$ is increased the numerical curves approach the result of the $A_{ij}$ theory for the original network. It is worth mentioning that the result of the $A_{ij}$ theory is the same for the original network and any of its $L$-cloned versions.

{\color{black}In the inset of Fig.~\subref*{fBa}, we illustrate the effect of $L$-cloning on two $\gamma(3,3)$ networks with different sizes. Since the curves in the inset that correspond to the same value of $L$ match each other, we conclude that the shift towards the $A_{ij}$ theory in the main Fig.~\subref*{fBa} with increasing $L$ is not due to the change in the network size, but because the density of triangles decreases for larger $L$. Moreover, the results of bond percolation on the $L_{-3}$-cloned versions of $\gamma(3,3)$ network (not shown) coincide with the numerical curve for the original network regardless of the number of layers. This is due to the fact that triangles are the only short loops present in $\gamma(3,3)$ networks and $L_{-3}$-cloning keeps triangles intact. Hence, $L_{-3}$-cloned $\gamma(3,3)$ networks have effectively the same structure as the original $\gamma(3,3)$ networks, i.e., triangles randomly connected via single edges (see Fig.~\ref{fg33}). These observations confirm that eliminating the short loops in the $\gamma(3,3)$ network, by $L$-cloning, transforms its structure to a treelike network for which the theory is designed, and that the difference between the theory and the numerics on the original $\gamma(3,3)$ network is mainly due to the triangles present in the network.
}

Figure~\subref*{fBb} shows the bond percolation results for the power grid network of the western United States \cite{Watts98} and its $L$-cloned versions. It has been previously observed that mean-field theories are very inaccurate for this network \cite{Melnik11,Gleeson12}, an example of which is $P_{k,k'}$ theory for bond percolation which employs Eqs.~(\ref{eqPkka}) and (\ref{eqPkkb}) \cite{Melnik11}. Figure~\subref*{fBb} shows that $A_{ij}$ theory is also inaccurate for this network; however, as $L$ is increased, the numerical result for the $L$-cloned version of the network approaches the prediction of the $A_{ij}$ theory. We conclude that the $A_{ij}$ theory for bond percolation will be accurate for \LC networks with sufficiently large $L$.

\begin{figure} [t] \centering
\includegraphics[width=1\columnwidth]{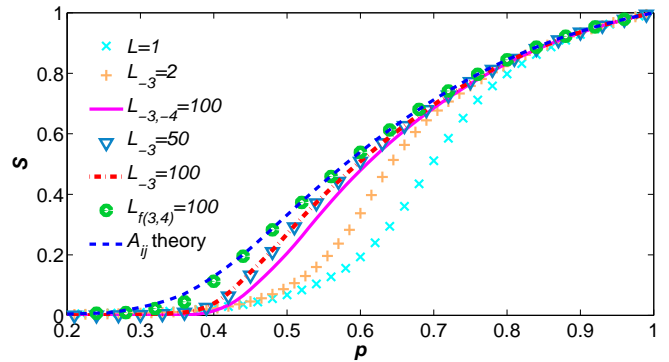}
\caption{\ed{(Color online) For a large number of layers, the simulation results of bond percolation on $L_{-3}$-cloned versions of the western US power grid \cite{Watts98} network converges to a curve that has an appreciable difference from the $A_{ij}$ theory. This indicates that the presence of triangles causes a significant error in the theoretical prediction. Similarly the numerics for $L_{-3,-4}$-cloned networks have an appreciable deviation from the theory; this deviation is even larger than that of $L_{-3}$-cloned networks, indicating that squares have also an appreciable effect on the accuracy of the theory. These deviations are not due to the smaller number of links rewired in $L_{-3}$ and $L_{-3,-4}$ cloned networks; since the numerics on $L_{f(3,4)}$ cloned networks (in which the fraction of rewired links is as low as that of $L_{-3,-4}$-cloned networks) with a large number of layers matches the theory very well. The number of layers is indicated by $L$ for each curve and the index of $L$ denotes the type of cloning.}} \label{fBP34}
\end{figure}

We also run numerical simulations of bond percolation on $L_{-3}$- and $L_{-3,-4}$-cloned versions of the power grid network. In Fig.~\ref{fBP34} we show that the difference between the $A_{ij}$ theory and numerics on $L_{-3}$-cloned networks decreases as we start to increase the number of layers $L_{-3}$. However, the numerical results on $L_{-3}$-cloned networks do not approach the $A_{ij}$ theory curve even for very large $L_{-3}$. Moreover, the numerical results on $L_{-3,-4}$-cloned networks differ from the $A_{ij}$ theory even more than the results on $L_{-3}$-cloned networks.

Due to the fact that in $L_{−3}$- and $L_{−3,−4}$-cloned networks short loops of specified lengths are preserved, to create these networks only the edges that are not part of such loops (hence only a fraction of the total number of edges) are reassigned (rewired), while in $L$-cloned networks all edges are rewired. For example, in the power grid network, 79\% of links are not part of loops of length 3 and 67\% of links are not part of loops of length 3 or 4. By running simulations on $L_{f}$-cloned networks, we show that if the same fraction of links, as in $L_{-3,-4}$-cloned networks, for example, are randomly selected and rewired between the layers, the numerical results are appreciably closer to the theoretical prediction, compared to the numerical results for the $L_{-3,-4}$-cloned networks. This implies that the preserved short loops, and not the smaller fraction of rewired links, is the reason that the numerics for $L_{-3,-4}$-cloned (and similarly for $L_{-3}$-cloned) networks do not match the $A_{ij}$ theory even for large number of layers.

In Fig.~\ref{fBP34} the numerical results for $L_{f(3,4)}$-cloned power grid are compared with the results for other variants of $L$-cloning. It is observed that the result for $L_{f}$-cloned networks, with a large number of layers $L_{f}$, matches very well with the $A_{ij}$ theory. This indicates that even with fewer rewired links, since the clustered structure is effectively destroyed for large $L_{f}$, the numerical results approach the adjacency tree-based theory. Similar results (not shown) were observed for site percolation.

These results demonstrate that triangles and also squares play a significant role in the inaccurate prediction of the $A_{ij}$ theory for the power grid network.
On the other hand, as the numerical results for $L_{-3,-4}$-cloned networks are closer to the theory than the numerics for the original network, there must exist other sources of error too. One source is the finite size effect,\footnote{{\color{black}
The finite size of the networks is known to cause deviations from the theoretical predictions; for example, the finite-size effect is shown to cause error in the $P_{k,k'}^{i,i'}$ theory for binary-state dynamics \cite{Melnik_Chaos2014}.}}
as the original network has a size smaller than any of its cloned versions. Other short loops with lengths larger than 4 may also have an effect on the accuracy of the $A_{ij}$ theory.

    \subsection{SI epidemic model}
\begin{figure} [b] \centering
\subfloat{\includegraphics[width=1\columnwidth]{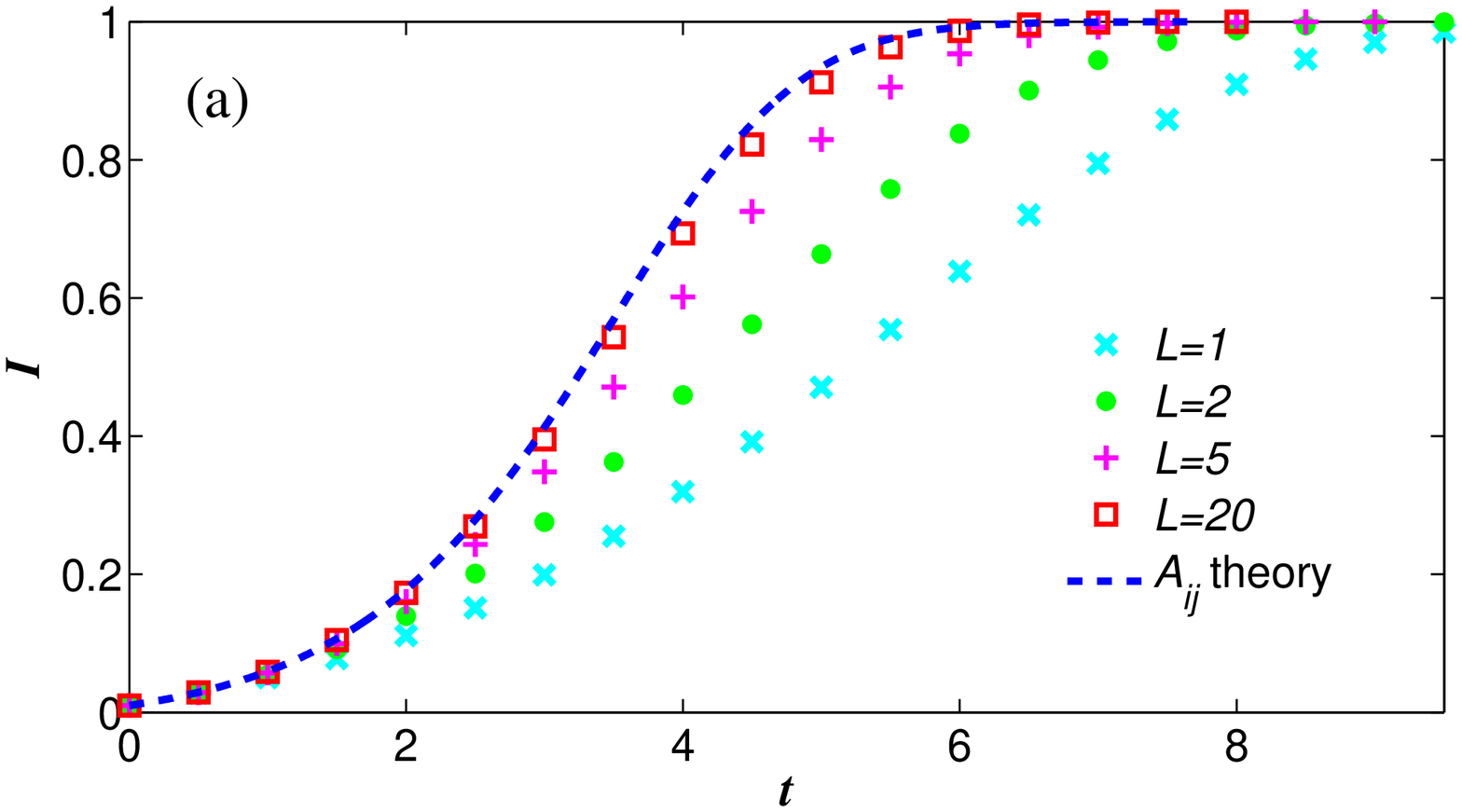}\label{fSa}}
\hfill
\subfloat{\includegraphics[width=1\columnwidth]{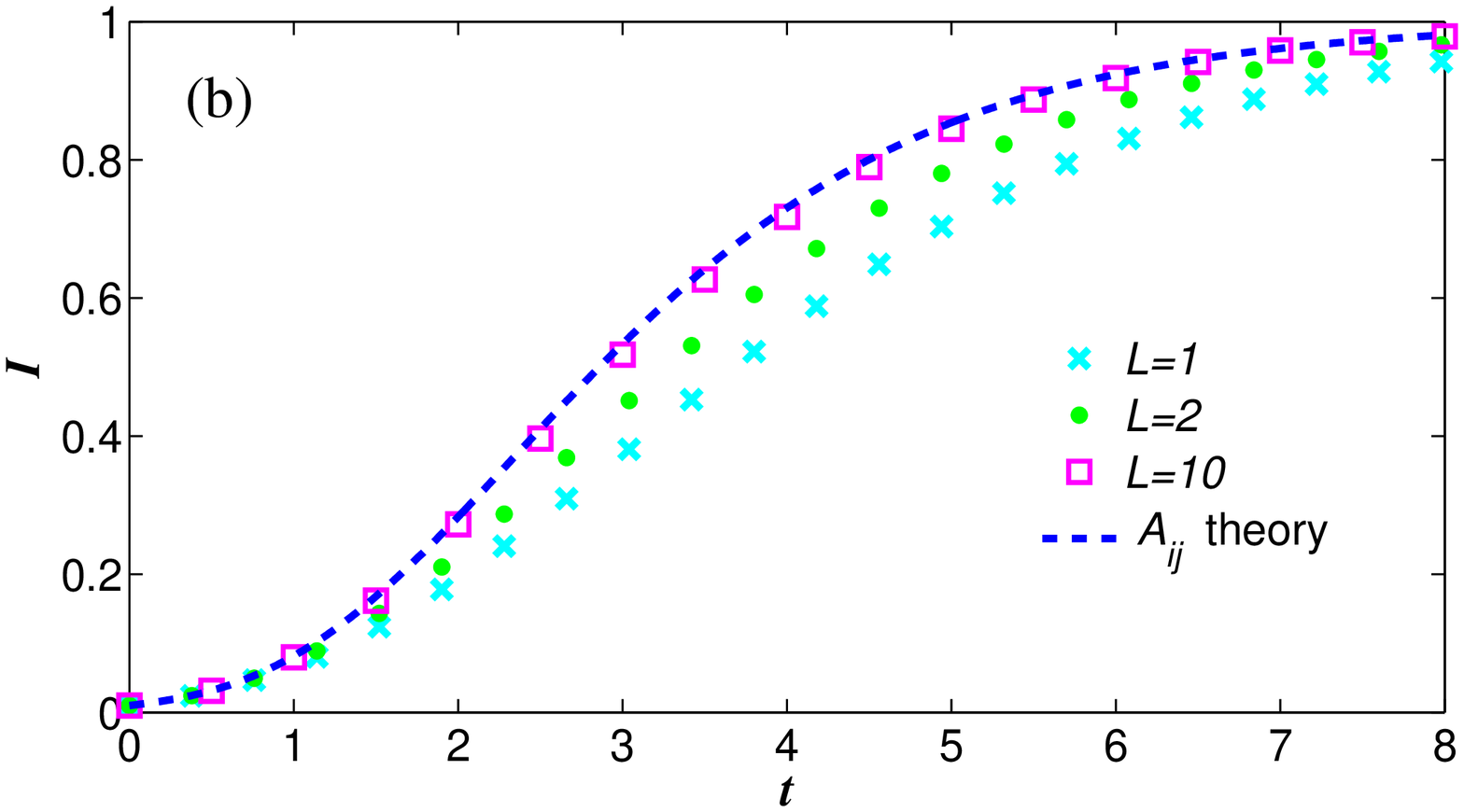}\label{fSb}}
\caption{(Color online) Numerical simulations and $A_{ij}$ theory for SI model on (a) a $\gamma(3,3)$ network with approximately $10^4$ nodes, and (b) the US power grid network and their corresponding $L$-cloned versions. Here $L=1$ indicates the results of numerical simulations on original networks. The $A_{ij}$ theory is not accurate on the original clustered networks, however its performance improves with $L$ on \LC networks.}
\label{fSI}
\end{figure}

In the Susceptible-Infected (SI) model of spread of disease, two possible states are considered for a node: It is either infected or susceptible \cite{Newman10}. The quantity of interest is the fraction of network nodes that are in the infected population $I$ at time $t$ after the disease begins spreading from a very small fraction of infected nodes. The probability $S_i$ that node $i$ is in the susceptible population at time $t$ can be obtained by using a pair approximation method and assuming that the network is treelike [see Eqs.~(17.54) and (17.55) of Ref. \cite{Newman10}]:
\begin{align}
\frac{dS_{i}}{dt} &= - \beta S_{i}\sum_{j} A_{ij}p_{ij} ~,\label{eS1}\\
\frac{dp_{ij}}{dt} &= \beta(1-p_{ij})[-p_{ij}+\sum_{k\neq i} A_{jk}p_{jk}] ~, \label{eS2}
\end{align}
where $p_{ij}$ is the conditional probability that node $j$ is infected given that node $i$ is susceptible. Then the fraction of the population that is infected is just $I=1-\sum_iS_i$. Since Eq.~\eqref{eS2} assumes that the network is locally treelike, we expect the predictions made by Eqs.~\eqref{eS1} and \eqref{eS2} to be inaccurate for clustered networks. Although there are cases where the theoretical results and actual behavior match quite well, such as in Fig.~(17.5) of Ref.~\cite{Newman10}, the accuracy of prediction of Eq.~\eqref{eS1} is not generally guaranteed for clustered networks. We demonstrate this in Fig.~\ref{fSI} by simulations of SI dynamics on the $\gamma(3,3)$ and the US power grid networks and their corresponding $L$-cloned versions: The theoretical predictions for these two clustered networks are only accurate for sufficiently large $L$.

{\color{black}
In order to examine the contribution of short loops of certain length against any other possible source of inaccuracy we employ the extensions of $L$-cloning introduced in Secs.~\ref{sec2d} and \ref{sec2e}. The numerical results of SI dynamics on $L_{-3}$-cloned versions of the $\gamma(3,3)$ networks (not shown) coincide with the results on the original network. This indicates that the presence of triangles is the source of inaccuracy of $A_{ij}$ theory for $\gamma(3,3)$ networks.
\begin{figure} [t] \centering
\includegraphics[width=1\columnwidth]{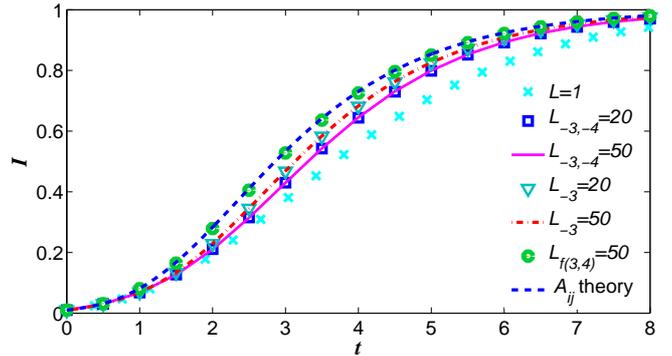}
\caption{\ed{(Color online) For a sufficiently large number of layers the results of numerical simulations for SI epidemics on $L_{-3}$ or $L_{-3,-4}$ cloned versions of the power grid network approach a specific curve, which is different from the $A_{ij}$ theory curve;
this is in contrast to the convergence of the results on $L$-cloned networks to the $A_{ij}$ theory. Although the fraction of reassigned (rewired) links in $L_{-3}$- or $L_{-3,-4}$ cloned networks is smaller than that of $L$-cloned networks, this is not the cause of difference in the corresponding numerical results, as the numerics on $L_{f(3,4)}$-cloned networks (in which the fraction of reassigned links is the same as that of the $L_{-3,-4}$-cloned network) with a large number of layers, match very well with the $A_{ij}$ theory.}} \label{fSI34}
\end{figure}

The numerics for the SI dynamics on $L_{-3}$- or $L_{-3,-4}$-cloned versions of the power grid network are closer to the $A_{ij}$ theory than those of the original network (Fig.\ref{fSI34}). However, the theory is still inaccurate on the $L_{-3}$- and $L_{-3,-4}$-cloned networks (even when the number of layers is large) because triangles or both triangles and squares are preserved in these networks. The difference between the numerics on $L$-cloned networks and the numerics on $L_{-3}$- or $L_{-3,-4}$-cloned networks is not due to the difference in the number of reassigned (rewired) links, because the numerics on $L_{f(3,4)}$-cloned networks (where the fraction of rewired links is the same as in $L_{-3,-4}$-cloned networks) matches the $A_{ij}$ theory very well (Fig.\ref{fSI34}). Similar to the results shown for bond percolation on the power grid network in Sec.~\ref{sec3a}, the SI numerics on $L_{-3}$- and $L_{-3,-4}$ cloned networks of the power grid are closer to the theoretical prediction; this may be due to finite size effects and the presence of loops of length greater than 4 in the original network. Nevertheless, the loops of lengths 3 and 4 are shown to have a significant effect on the accuracy of the $A_{ij}$ theory for SI dynamics.
}

    \subsection{The Ising model}
The Ising model is a simplified theoretical framework describing the local interactions between magnetic moments of atomic or multi-atomic particles in a solid \cite{kittel}. In the Ising model each node can be in one of the two spin states called up or down (here denoted by $+1$ and $-1$), and the spin of each node is affected by the spins of its neighbors in the network through the Ising Hamiltonian \cite{kittel,Dorogovtsev08}. This Hamiltonian determines the equilibrium configuration of spin states in the network. The total magnetic moment $M$ of the system at equilibrium is defined as the sum of local spins. For a locally treelike network the expected spin of nodes at equilibrium according to belief propagation algorithm \cite{Dorogovtsev08} is given by the set of equations described in Sec. VI.A.2 of Ref.~\cite{Dorogovtsev08_arxiv}. For constant coupling $J$ and in the absence of external magnetic field\footnote{These assumptions are not necessary, however we consider the simplest case which makes the equations shorter and simpler, but still supports the argument.} these equations are:
\begin{eqnarray}
  &\mu_{ji}(S_i)&=R\sum_{S_j=\pm 1}e^{\beta JS_iS_j}\prod_{n\in N(j)\backslash i} \mu_{nj}(S_j) ~, \label{eqIs1}\\
  &b_i(S_i)&= R\prod_{j~\in N(i)}\mu_{ji}(S_i) ~, \label{eqIs2}\\
  &M_i&=\sum_{S_i=\pm1}S_ib_i(S_i) ~ \label{eqIs3},
\end{eqnarray}
where $\beta=1/(kT)$ with $k$ being the Boltzmann constant, $T$ is the temperature, $R$ is a normalization constant, $S_i$ is the spin value at node $i$, and $\mu_{ij}$ are called \emph{messages} in the belief propagation method. The product in Eq.~(\ref{eqIs1}) is over all neighbors of node $j$ except $i$. The fixed point of Eq.~(\ref{eqIs1}) is used to calculate $b_i(S_i)$, the probability that node $i$ is in state $S_i$ at equilibrium. Accordingly, the expected local magnetic momentum $M_i$ is calculated from Eq.~(\ref{eqIs3}). The result of the belief propagation method for the Ising model is equivalent to that of the Bethe-Peierls approach \cite{Dorogovtsev08_arxiv}. This can be shown by writing $\mu_{ij}$ in a general form as:
\begin{equation}\label{eqIs4}
  \mu_{ji}(S_i)=\frac{e^{\beta h_{ji}S_i}}{2\cosh\beta h_{ji}},
\end{equation}
which transforms Eqs.~(\ref{eqIs1})$-$(\ref{eqIs3}) to the Bethe-Peierls equations of Ref. \cite{Dorogovtsev08_arxiv}.

\begin{figure} [t] \centering
\includegraphics[width=1\columnwidth]{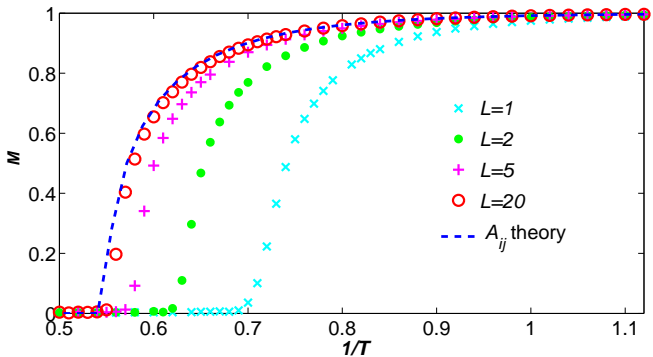}
\caption{(Color online) Magnetization $M$ versus inverse temperature for the Ising model on a $\gamma(3,3)$ network and its \LC versions. Here $L=1$ indicates the result of numerical simulations on the original $\gamma(3,3)$ network with approximately $10^6$ nodes, averaged over 20 realizations of the Ising model. The $A_{ij}$ theory is inaccurate on this network; however its performance improves appreciably with $L$ on the \LC versions of this network.} \label{fI}
\end{figure}

These two methods are $A_{ij}$ theories; hence, as in the previous examples, their predictions for clustered networks are prone to errors. Figure~\ref{fI} illustrates the behavior of the magnetization $M$ versus inverse temperature $1/T$ (for $J=1$ and $k=1$) in a $\gamma(3,3)$ network and its $L$-cloned versions; the results are compared to the theoretical prediction from the above equations. As expected, the tree-based approach does not show the correct behavior for the original $\gamma(3,3)$ network. However, for sufficiently large $L$, the prediction of $A_{ij}$ theories matches the numerical simulations on the corresponding \LC versions of the network.
{\color{black}
It is worth mentioning that, as expected, the numerical results on $L_{-3}$-cloned $\gamma(3,3)$ network (not shown) match the numerics on the original $\gamma(3,3)$ network. This implies that the improved agreement between the numerical results on $L$-cloned networks and the $A_{ij}$ theory is not due to the larger size of $L$-cloned networks. Hence, the presence of triangles is the source of inaccuracy in the $A_{ij}$ theory for the Ising model on this network.
}

\section{Conclusion} \label{sec4}
We introduced the so-called $L$-cloning of networks, a new technique to create random ensembles of networks with certain properties based on real-world or synthetic complex networks. We demonstrated in Sec.~\ref{sec2} that \LC networks have degree distribution and degree-degree correlation between and beyond nearest neighbors that are identical to those in the original network from which they are constructed. However, the density of triangles in $L$-cloned networks, and hence the clustering coefficient $C$, is reduced by a factor of $L$. Moreover, the expected density of any short loops approaches zero for sufficiently large $L$.
{\color{black}
Some useful extensions of \LC networks, i.e., $L_{-3}$-, $L_{-3,-4}$-, and $L_f$-cloned networks, were also introduced; similar to \LC networks, these networks have degree distribution and degree-degree correlations between and beyond nearest neighbors identical to those of the original network from which they are constructed. On the other hand, in $L_{-3}$-cloned networks the triangles are preserved, and in $L_{-3,-4}$-cloned networks both triangles and squares are preserved.

We used these properties of network cloning to investigate the effect of short loops on dynamical processes running on networks and to inspect the applicability of tree-based theories to clustered networks (i.e., networks with appreciable density of short loops). In this regard the accuracy of theories for percolation, SI epidemics, and the Ising model were investigated by comparing the theoretical predictions against numerical simulations on examples of clustered networks and their cloned versions.

We demonstrated that \LC networks with sufficiently large $L$ are the ensemble of networks for which the $A_{ij}$ theories are designed. Hence, by running numerical simulations on L-cloned networks one can predict the outcome and potential benefits of $A_{ij}$ theories for dynamical processes where such theories do not yet exist. In addition, by comparing the numerics for the dynamics on $L_{-3}$ and $L_{-3,-4}$-cloned networks with the numerics on \LC networks we highlighted the effect of triangles and squares on the deviation of numerical results from the theoretical predictions. Using $L_f$-cloned networks we ruled out any possible effect due to the different fraction of reassigned (rewired) links in \LC networks and in $L_{-3}$- or $L_{-3,-4}$-cloned networks and demonstrated the significant effect of short loops on the accuracy of the $A_{ij}$ theories.}

Nevertheless, it was previously shown~\cite{Gleeson12} that mean-field theories can perform well on clustered networks with high values of mean nearest-neighbor degree. Moreover, the existence of a double percolation phase transition on some clustered networks has recently been reported \cite{colomer2014double}, an effect that is not considered in the tree-based theories. These findings imply that although network clustering can cause inaccuracy in tree-based theories, its net effect and strength depend also on other factors. In this regard, inspecting the organization of clustered structures in networks (e.g., as in Ref. \cite{colomer2013deciphering}) can contribute to the understanding of the effect of network clustering on dynamical processes, and its mitigation by, for example, high values of mean nearest-neighbor degree.

It is worth noting that the belief propagation and Bethe-Peierls methods used for deriving analytical results for the Ising model have also been employed in many other problems in various areas \cite{Dorogovtsev08}. According to the tree-based nature of these approximations, the models for which they are used are subject to inaccuracies due to network clustering (short loops). This suggests that network $L$-cloning and its extensions can be applied to gain a better understanding of the accuracy of these models as well.

Another potential application of cloned networks is the analysis of the finite size effect on real systems, where a small sample of a large network is available. By inspecting the result of numerical simulations on the sample and on its \LC versions, one can examine the sensitivity of results to changes in network size. If the network of interest is clustered, the changes made in network clustering should be considered in addition to the finite size effect. {\color{black} The extensions of $L$-cloning that preserve the density of loops of specified lengths can help achieve this goal.
}

\section*{Acknowledgements}
A.F., S.M., and J.P.G. acknowledge funding provided by Science Foundation Ireland (SFI) under programme 11/PI/1026. J.P.G. acknowledges funding  provided by the European Commission FET-Proactive project PLEXMATH (FP7-ICT-2011-8; Grant No. 317614), and by Science Foundation Ireland under programme 12/IA/1683. We thank Adam Hackett for helpful discussions, and acknowledge the SFI/HEA Irish Centre for High-End Computing (ICHEC) for the provision of computational facilities.

\section*{Appendix: Relation between $L$-cloned and multilayer networks}\label{Apx}
To describe an $L$-cloned network as a multilayer network, we use the terminology of Ref.~\cite{kivela2014JCN}. Accordingly, an \LC network has one \emph{aspect} on which there exist $L$ layers of the network and all the layers have an equal size. The inter-layer connections are not \emph{couplings}, hence $L$-cloned networks does not have any of the \emph{diagonal}, \emph{layer-coupled} or \emph{categorial} couplings. If the clones of each node are assumed to be the same node then \LC networks are \emph{node-aligned}; otherwise if different identities are assumed for clones of each node, then \LC networks are \emph{layer-disjoint} networks. The former can be argued according to the fact that all clones of a certain node have the same degree and degree-degree correlation between and beyond nearest neighbors (i.e., the degree sequence $\{K\}_{(i,d)}$ of nodes at any distance $d$ from any node $i$ in an $L$-cloned network is identical to that of all other clones of node $i$). However, as other structural characteristics of clones of a node (e.g., closeness and betweenness~\cite{Newman10}) are not exactly the same, assuming $L$-cloned networks to be the latter case, i.e., \emph{layer-disjoint} networks, is also legitimate and informative.

\bibliography{networks}

\begin{thebibliography}{26}%
\makeatletter
\providecommand \@ifxundefined [1]{%
 \@ifx{#1\undefined}
}%
\providecommand \@ifnum [1]{%
 \ifnum #1\expandafter \@firstoftwo
 \else \expandafter \@secondoftwo
 \fi
}%
\providecommand \@ifx [1]{%
 \ifx #1\expandafter \@firstoftwo
 \else \expandafter \@secondoftwo
 \fi
}%
\providecommand \natexlab [1]{#1}%
\providecommand \enquote  [1]{``#1''}%
\providecommand \bibnamefont  [1]{#1}%
\providecommand \bibfnamefont [1]{#1}%
\providecommand \citenamefont [1]{#1}%
\providecommand \href@noop [0]{\@secondoftwo}%
\providecommand \href [0]{\begingroup \@sanitize@url \@href}%
\providecommand \@href[1]{\@@startlink{#1}\@@href}%
\providecommand \@@href[1]{\endgroup#1\@@endlink}%
\providecommand \@sanitize@url [0]{\catcode `\\12\catcode `\$12\catcode
  `\&12\catcode `\#12\catcode `\^12\catcode `\_12\catcode `\%12\relax}%
\providecommand \@@startlink[1]{}%
\providecommand \@@endlink[0]{}%
\providecommand \url  [0]{\begingroup\@sanitize@url \@url }%
\providecommand \@url [1]{\endgroup\@href {#1}{\urlprefix }}%
\providecommand \urlprefix  [0]{URL }%
\providecommand \Eprint [0]{\href }%
\providecommand \doibase [0]{http://dx.doi.org/}%
\providecommand \selectlanguage [0]{\@gobble}%
\providecommand \bibinfo  [0]{\@secondoftwo}%
\providecommand \bibfield  [0]{\@secondoftwo}%
\providecommand \translation [1]{[#1]}%
\providecommand \BibitemOpen [0]{}%
\providecommand \bibitemStop [0]{}%
\providecommand \bibitemNoStop [0]{.\EOS\space}%
\providecommand \EOS [0]{\spacefactor3000\relax}%
\providecommand \BibitemShut  [1]{\csname bibitem#1\endcsname}%
\let\auto@bib@innerbib\@empty
%</preamble>
\bibitem [{\citenamefont {Newman}(2010)}]{Newman10}%
  \BibitemOpen
  \bibfield  {author} {\bibinfo {author} {\bibfnamefont {M.~E.~J.}\
  \bibnamefont {Newman}},\ }\href@noop {} {\emph {\bibinfo {title} {Networks:
  An Introduction}}}\ (\bibinfo  {publisher} {Oxford University Press},\
  \bibinfo {address} {Oxford},\ \bibinfo {year} {2010})\BibitemShut {NoStop}%
\bibitem [{\citenamefont {Newman}(2009)}]{Newman09}%
  \BibitemOpen
  \bibfield  {author} {\bibinfo {author} {\bibfnamefont {M.~E.~J.}\
  \bibnamefont {Newman}},\ }\bibfield  {title} {\enquote {\bibinfo {title}
  {Random graphs with clustering},}\ }\href@noop {} {\bibfield  {journal}
  {\bibinfo  {journal} {Phys. Rev. Lett.}\ }\textbf {\bibinfo {volume} {103}},\
  \bibinfo {pages} {058701} (\bibinfo {year} {2009})}\BibitemShut {NoStop}%
\bibitem [{\citenamefont {Miller}(2009)}]{Miller09b}%
  \BibitemOpen
  \bibfield  {author} {\bibinfo {author} {\bibfnamefont {J.~C.}\ \bibnamefont
  {Miller}},\ }\bibfield  {title} {\enquote {\bibinfo {title} {Percolation and
  epidemics in random clustered networks},}\ }\href@noop {} {\bibfield
  {journal} {\bibinfo  {journal} {Phys. Rev. E}\ }\textbf {\bibinfo {volume}
  {80}},\ \bibinfo {pages} {020901(R)} (\bibinfo {year} {2009})}\BibitemShut
  {NoStop}%
\bibitem [{\citenamefont {Gleeson}(2009)}]{Gleeson09a}%
  \BibitemOpen
  \bibfield  {author} {\bibinfo {author} {\bibfnamefont {J.~P.}\ \bibnamefont
  {Gleeson}},\ }\bibfield  {title} {\enquote {\bibinfo {title} {Bond
  percolation on a class of clustered random networks},}\ }\href@noop {}
  {\bibfield  {journal} {\bibinfo  {journal} {Phys. Rev. E}\ }\textbf {\bibinfo
  {volume} {80}},\ \bibinfo {pages} {036107} (\bibinfo {year}
  {2009})}\BibitemShut {NoStop}%
\bibitem [{\citenamefont {Boccaletti}\ \emph {et~al.}(2006)\citenamefont
  {Boccaletti}, \citenamefont {Latora}, \citenamefont {Moreno}, \citenamefont
  {Chavez},\ and\ \citenamefont {Hwang}}]{Boccaletti06}%
  \BibitemOpen
  \bibfield  {author} {\bibinfo {author} {\bibfnamefont {S.}~\bibnamefont
  {Boccaletti}}, \bibinfo {author} {\bibfnamefont {V.}~\bibnamefont {Latora}},
  \bibinfo {author} {\bibfnamefont {Y.}~\bibnamefont {Moreno}}, \bibinfo
  {author} {\bibfnamefont {M.}~\bibnamefont {Chavez}}, \ and\ \bibinfo {author}
  {\bibfnamefont {D.-U.}\ \bibnamefont {Hwang}},\ }\bibfield  {title} {\enquote
  {\bibinfo {title} {Complex networks: Structure and dynamics},}\ }\href@noop
  {} {\bibfield  {journal} {\bibinfo  {journal} {Phys. Rep.}\ }\textbf
  {\bibinfo {volume} {424}},\ \bibinfo {pages} {175} (\bibinfo {year}
  {2006})}\BibitemShut {NoStop}%
\bibitem [{\citenamefont {Newman}(2003)}]{Newman03a}%
  \BibitemOpen
  \bibfield  {author} {\bibinfo {author} {\bibfnamefont {M.~E.~J.}\
  \bibnamefont {Newman}},\ }\bibfield  {title} {\enquote {\bibinfo {title} {The
  structure and function of complex networks},}\ }\href@noop {} {\bibfield
  {journal} {\bibinfo  {journal} {SIAM Rev.}\ }\textbf {\bibinfo {volume}
  {45}},\ \bibinfo {pages} {167} (\bibinfo {year} {2003})}\BibitemShut
  {NoStop}%
\bibitem [{\citenamefont {Karrer}\ and\ \citenamefont
  {Newman}(2010{\natexlab{a}})}]{Karrer82}%
  \BibitemOpen
  \bibfield  {author} {\bibinfo {author} {\bibfnamefont {B.}~\bibnamefont
  {Karrer}}\ and\ \bibinfo {author} {\bibfnamefont {M.~E.~J.}\ \bibnamefont
  {Newman}},\ }\bibfield  {title} {\enquote {\bibinfo {title} {Random graphs
  containing arbitrary distributions of subgraphs},}\ }\href {\doibase
  10.1103/PhysRevE.82.066118} {\bibfield  {journal} {\bibinfo  {journal} {Phys.
  Rev. E}\ }\textbf {\bibinfo {volume} {82}},\ \bibinfo {pages} {066118}
  (\bibinfo {year} {2010}{\natexlab{a}})}\BibitemShut {NoStop}%
\bibitem [{\citenamefont {Gleeson}\ and\ \citenamefont
  {Melnik}(2009)}]{Gleeson09b}%
  \BibitemOpen
  \bibfield  {author} {\bibinfo {author} {\bibfnamefont {J.~P.}\ \bibnamefont
  {Gleeson}}\ and\ \bibinfo {author} {\bibfnamefont {S.}~\bibnamefont
  {Melnik}},\ }\bibfield  {title} {\enquote {\bibinfo {title} {Analytical
  results for bond percolation and k-core sizes on clustered networks},}\
  }\href@noop {} {\bibfield  {journal} {\bibinfo  {journal} {Phys. Rev. E}\
  }\textbf {\bibinfo {volume} {80}},\ \bibinfo {pages} {046121} (\bibinfo
  {year} {2009})}\BibitemShut {NoStop}%
\bibitem [{\citenamefont {Hackett}\ and\ \citenamefont
  {Gleeson}(2013)}]{Hackett12}%
  \BibitemOpen
  \bibfield  {author} {\bibinfo {author} {\bibfnamefont {A.}~\bibnamefont
  {Hackett}}\ and\ \bibinfo {author} {\bibfnamefont {J.~P.}\ \bibnamefont
  {Gleeson}},\ }\bibfield  {title} {\enquote {\bibinfo {title} {Cascades on
  clique-based graphs},}\ }\href {\doibase 10.1103/PhysRevE.87.062801}
  {\bibfield  {journal} {\bibinfo  {journal} {Phys. Rev. E}\ }\textbf {\bibinfo
  {volume} {87}},\ \bibinfo {pages} {062801} (\bibinfo {year}
  {2013})}\BibitemShut {NoStop}%
\bibitem [{\citenamefont {Melnik}\ \emph {et~al.}(2011)\citenamefont {Melnik},
  \citenamefont {Hackett}, \citenamefont {Porter}, \citenamefont {Mucha},\ and\
  \citenamefont {Gleeson}}]{Melnik11}%
  \BibitemOpen
  \bibfield  {author} {\bibinfo {author} {\bibfnamefont {S.}~\bibnamefont
  {Melnik}}, \bibinfo {author} {\bibfnamefont {A.}~\bibnamefont {Hackett}},
  \bibinfo {author} {\bibfnamefont {M.~A.}\ \bibnamefont {Porter}}, \bibinfo
  {author} {\bibfnamefont {P.~J.}\ \bibnamefont {Mucha}}, \ and\ \bibinfo
  {author} {\bibfnamefont {J.~P.}\ \bibnamefont {Gleeson}},\ }\bibfield
  {title} {\enquote {\bibinfo {title} {The unreasonable effectiveness of
  tree-based theory for networks with clustering},}\ }\href@noop {} {\bibfield
  {journal} {\bibinfo  {journal} {Phys. Rev. E}\ }\textbf {\bibinfo {volume}
  {83}},\ \bibinfo {pages} {036112} (\bibinfo {year} {2011})}\BibitemShut
  {NoStop}%
\bibitem [{\citenamefont {Gleeson}\ \emph {et~al.}(2012)\citenamefont
  {Gleeson}, \citenamefont {Melnik}, \citenamefont {Ward}, \citenamefont
  {Porter},\ and\ \citenamefont {Mucha}}]{Gleeson12}%
  \BibitemOpen
  \bibfield  {author} {\bibinfo {author} {\bibfnamefont {J.~P.}\ \bibnamefont
  {Gleeson}}, \bibinfo {author} {\bibfnamefont {S.}~\bibnamefont {Melnik}},
  \bibinfo {author} {\bibfnamefont {J.~A.}\ \bibnamefont {Ward}}, \bibinfo
  {author} {\bibfnamefont {M.~A.}\ \bibnamefont {Porter}}, \ and\ \bibinfo
  {author} {\bibfnamefont {P.~J.}\ \bibnamefont {Mucha}},\ }\bibfield  {title}
  {\enquote {\bibinfo {title} {Accuracy of mean-field theory for dynamics on
  real-world networks},}\ }\href@noop {} {\bibfield  {journal} {\bibinfo
  {journal} {Phys. Rev. E}\ }\textbf {\bibinfo {volume} {85}},\ \bibinfo
  {pages} {026106} (\bibinfo {year} {2012})}\BibitemShut {NoStop}%
\bibitem [{\citenamefont {Gleeson}(2013)}]{Gleeson13}%
  \BibitemOpen
  \bibfield  {author} {\bibinfo {author} {\bibfnamefont {J.~P.}\ \bibnamefont
  {Gleeson}},\ }\bibfield  {title} {\enquote {\bibinfo {title} {Binary-state
  dynamics on complex networks: Pair approximation and beyond},}\ }\href
  {\doibase 10.1103/PhysRevX.3.021004} {\bibfield  {journal} {\bibinfo
  {journal} {Phys. Rev. X}\ }\textbf {\bibinfo {volume} {3}},\ \bibinfo {pages}
  {021004} (\bibinfo {year} {2013})}\BibitemShut {NoStop}%
\bibitem [{\citenamefont {Restrepo}\ \emph {et~al.}(2008)\citenamefont
  {Restrepo}, \citenamefont {Ott},\ and\ \citenamefont
  {Hunt}}]{restrepo2008weighted}%
  \BibitemOpen
  \bibfield  {author} {\bibinfo {author} {\bibfnamefont {J.~G.}\ \bibnamefont
  {Restrepo}}, \bibinfo {author} {\bibfnamefont {E.}~\bibnamefont {Ott}}, \
  and\ \bibinfo {author} {\bibfnamefont {B.~R.}\ \bibnamefont {Hunt}},\
  }\bibfield  {title} {\enquote {\bibinfo {title} {Weighted percolation on
  directed networks},}\ }\href@noop {} {\bibfield  {journal} {\bibinfo
  {journal} {Phys. Rev. Lett.}\ }\textbf {\bibinfo {volume} {100}},\ \bibinfo
  {pages} {058701} (\bibinfo {year} {2008})}\BibitemShut {NoStop}%
\bibitem [{\citenamefont {Karrer}\ \emph {et~al.}(2014)\citenamefont {Karrer},
  \citenamefont {Newman},\ and\ \citenamefont {Zdeborov\'a}}]{karrer_PRL14}%
  \BibitemOpen
  \bibfield  {author} {\bibinfo {author} {\bibfnamefont {B.}~\bibnamefont
  {Karrer}}, \bibinfo {author} {\bibfnamefont {M.~E.~J.}\ \bibnamefont
  {Newman}}, \ and\ \bibinfo {author} {\bibfnamefont {L.}~\bibnamefont
  {Zdeborov\'a}},\ }\bibfield  {title} {\enquote {\bibinfo {title} {Percolation
  on sparse networks},}\ }\href {\doibase 10.1103/PhysRevLett.113.208702}
  {\bibfield  {journal} {\bibinfo  {journal} {Phys. Rev. Lett.}\ }\textbf
  {\bibinfo {volume} {113}},\ \bibinfo {pages} {208702} (\bibinfo {year}
  {2014})}\BibitemShut {NoStop}%
\bibitem [{\citenamefont {Dorogovtsev}\ \emph {et~al.}(2008)\citenamefont
  {Dorogovtsev}, \citenamefont {Goltsev},\ and\ \citenamefont
  {Mendes}}]{Dorogovtsev08}%
  \BibitemOpen
  \bibfield  {author} {\bibinfo {author} {\bibfnamefont {S.~N.}\ \bibnamefont
  {Dorogovtsev}}, \bibinfo {author} {\bibfnamefont {A.~V.}\ \bibnamefont
  {Goltsev}}, \ and\ \bibinfo {author} {\bibfnamefont {J.~F.~F.}\ \bibnamefont
  {Mendes}},\ }\bibfield  {title} {\enquote {\bibinfo {title} {Critical
  phenomena in complex networks},}\ }\href@noop {} {\bibfield  {journal}
  {\bibinfo  {journal} {Rev. Mod. Phys.}\ }\textbf {\bibinfo {volume} {80}},\
  \bibinfo {pages} {1275} (\bibinfo {year} {2008})}\BibitemShut {NoStop}%
\bibitem [{\citenamefont {Shrestha}\ and\ \citenamefont
  {Moore}(2014)}]{Shrestha_PRE14}%
  \BibitemOpen
  \bibfield  {author} {\bibinfo {author} {\bibfnamefont {M.}~\bibnamefont
  {Shrestha}}\ and\ \bibinfo {author} {\bibfnamefont {C.}~\bibnamefont
  {Moore}},\ }\bibfield  {title} {\enquote {\bibinfo {title} {Message-passing
  approach for threshold models of behavior in networks},}\ }\href {\doibase
  10.1103/PhysRevE.89.022805} {\bibfield  {journal} {\bibinfo  {journal} {Phys.
  Rev. E}\ }\textbf {\bibinfo {volume} {89}},\ \bibinfo {pages} {022805}
  (\bibinfo {year} {2014})}\BibitemShut {NoStop}%
\bibitem [{\citenamefont {Kivel{\"a}}\ \emph {et~al.}(2014)\citenamefont
  {Kivel{\"a}}, \citenamefont {Arenas}, \citenamefont {Barthelemy},
  \citenamefont {Gleeson}, \citenamefont {Moreno},\ and\ \citenamefont
  {Porter}}]{kivela2014JCN}%
  \BibitemOpen
  \bibfield  {author} {\bibinfo {author} {\bibfnamefont {M.}~\bibnamefont
  {Kivel{\"a}}}, \bibinfo {author} {\bibfnamefont {A.}~\bibnamefont {Arenas}},
  \bibinfo {author} {\bibfnamefont {M.}~\bibnamefont {Barthelemy}}, \bibinfo
  {author} {\bibfnamefont {J.~P.}\ \bibnamefont {Gleeson}}, \bibinfo {author}
  {\bibfnamefont {Y.}~\bibnamefont {Moreno}}, \ and\ \bibinfo {author}
  {\bibfnamefont {M.~A.}\ \bibnamefont {Porter}},\ }\bibfield  {title}
  {\enquote {\bibinfo {title} {Multilayer networks},}\ }\href@noop {}
  {\bibfield  {journal} {\bibinfo  {journal} {Journal of Complex Networks}\
  }\textbf {\bibinfo {volume} {2}},\ \bibinfo {pages} {203} (\bibinfo {year}
  {2014})}\BibitemShut {NoStop}%
\bibitem [{\citenamefont {Watts}\ and\ \citenamefont
  {Strogatz}(1998)}]{Watts98}%
  \BibitemOpen
  \bibfield  {author} {\bibinfo {author} {\bibfnamefont {D.~J.}\ \bibnamefont
  {Watts}}\ and\ \bibinfo {author} {\bibfnamefont {S.~H.}\ \bibnamefont
  {Strogatz}},\ }\bibfield  {title} {\enquote {\bibinfo {title} {Collective
  dynamics of 'small-world' networks},}\ }\href@noop {} {\bibfield  {journal}
  {\bibinfo  {journal} {Nature (London)}\ }\textbf {\bibinfo {volume} {393}},\
  \bibinfo {pages} {440} (\bibinfo {year} {1998})}\BibitemShut {NoStop}%
\bibitem [{\citenamefont {Karrer}\ and\ \citenamefont
  {Newman}(2010{\natexlab{b}})}]{Karrer10}%
  \BibitemOpen
  \bibfield  {author} {\bibinfo {author} {\bibfnamefont {B.}~\bibnamefont
  {Karrer}}\ and\ \bibinfo {author} {\bibfnamefont {M.~E.~J.}\ \bibnamefont
  {Newman}},\ }\bibfield  {title} {\enquote {\bibinfo {title} {Message passing
  approach for general epidemic models},}\ }\href@noop {} {\bibfield  {journal}
  {\bibinfo  {journal} {Phys. Rev. E}\ }\textbf {\bibinfo {volume} {82}},\
  \bibinfo {pages} {016101} (\bibinfo {year} {2010}{\natexlab{b}})}\BibitemShut
  {NoStop}%
\bibitem [{\citenamefont {Porter}\ and\ \citenamefont
  {Gleeson}()}]{porter2014dynamical}%
  \BibitemOpen
  \bibfield  {author} {\bibinfo {author} {\bibfnamefont {M.~A.}\ \bibnamefont
  {Porter}}\ and\ \bibinfo {author} {\bibfnamefont {J.~P.}\ \bibnamefont
  {Gleeson}},\ }\bibfield  {title} {\enquote {\bibinfo {title} {Dynamical
  systems on networks: A tutorial},}\ }\href@noop {} {\bibinfo  {journal}
  {arXiv:1403.7663}\ }\BibitemShut {NoStop}%
\bibitem [{\citenamefont {V\'{a}zquez}\ and\ \citenamefont
  {Moreno}(2003)}]{Vazquez03}%
  \BibitemOpen
\bibfield  {journal} {  }\bibfield  {author} {\bibinfo {author} {\bibfnamefont
  {A.}~\bibnamefont {V\'{a}zquez}}\ and\ \bibinfo {author} {\bibfnamefont
  {Y.}~\bibnamefont {Moreno}},\ }\bibfield  {title} {\enquote {\bibinfo {title}
  {Resilience to damage of graphs with degree correlations},}\ }\href@noop {}
  {\bibfield  {journal} {\bibinfo  {journal} {Phys. Rev. E}\ }\textbf {\bibinfo
  {volume} {67}},\ \bibinfo {pages} {015101(R)} (\bibinfo {year}
  {2003})}\BibitemShut {NoStop}%
\bibitem [{\citenamefont {Melnik}\ \emph {et~al.}(2014)\citenamefont {Melnik},
  \citenamefont {Porter}, \citenamefont {Mucha},\ and\ \citenamefont
  {Gleeson}}]{Melnik_Chaos2014}%
  \BibitemOpen
  \bibfield  {author} {\bibinfo {author} {\bibfnamefont {S.}~\bibnamefont
  {Melnik}}, \bibinfo {author} {\bibfnamefont {M.~A.}\ \bibnamefont {Porter}},
  \bibinfo {author} {\bibfnamefont {P.~J.}\ \bibnamefont {Mucha}}, \ and\
  \bibinfo {author} {\bibfnamefont {J.~P.}\ \bibnamefont {Gleeson}},\
  }\bibfield  {title} {\enquote {\bibinfo {title} {Dynamics on modular networks
  with heterogeneous correlations},}\ }\href@noop {} {\bibfield  {journal}
  {\bibinfo  {journal} {Chaos}\ }\textbf {\bibinfo {volume} {24}},\ \bibinfo
  {pages} {023106} (\bibinfo {year} {2014})}\BibitemShut {NoStop}%
\bibitem [{\citenamefont {Kittel}(2004)}]{kittel}%
  \BibitemOpen
  \bibfield  {author} {\bibinfo {author} {\bibfnamefont {C.}~\bibnamefont
  {Kittel}},\ }\href {http://books.google.ie/books?id=kym4QgAACAAJ} {\emph
  {\bibinfo {title} {Introduction to Solid State Physics}}}\ (\bibinfo
  {publisher} {Wiley, New York},\ \bibinfo {year} {2004})\BibitemShut {NoStop}%
\bibitem [{\citenamefont {Dorogovtsev}\ \emph {et~al.}()\citenamefont
  {Dorogovtsev}, \citenamefont {Goltsev},\ and\ \citenamefont
  {Mendes}}]{Dorogovtsev08_arxiv}%
  \BibitemOpen
  \bibfield  {author} {\bibinfo {author} {\bibfnamefont {S.~N.}\ \bibnamefont
  {Dorogovtsev}}, \bibinfo {author} {\bibfnamefont {A.~V.}\ \bibnamefont
  {Goltsev}}, \ and\ \bibinfo {author} {\bibfnamefont {J.~F.~F.}\ \bibnamefont
  {Mendes}},\ }\bibfield  {title} {\enquote {\bibinfo {title} {Critical
  phenomena in complex networks},}\ }\href {http://arxiv.org/abs/0705.0010}
  {\bibinfo  {journal} {arXiv:0705.0010v6}\ }\BibitemShut {NoStop}%
\bibitem [{\citenamefont {Colomer{-}de{-}Sim\'on}\ and\ \citenamefont
  {Bogu\~n\'a}(2014)}]{colomer2014double}%
  \BibitemOpen
\bibfield  {journal} {  }\bibfield  {author} {\bibinfo {author} {\bibfnamefont
  {P.}~\bibnamefont {Colomer{-}de{-}Sim\'on}}\ and\ \bibinfo {author}
  {\bibfnamefont {M.}~\bibnamefont {Bogu\~n\'a}},\ }\bibfield  {title}
  {\enquote {\bibinfo {title} {Double percolation phase transition in clustered
  complex networks},}\ }\href {\doibase 10.1103/PhysRevX.4.041020} {\bibfield
  {journal} {\bibinfo  {journal} {Phys. Rev. X}\ }\textbf {\bibinfo {volume}
  {4}},\ \bibinfo {pages} {041020} (\bibinfo {year} {2014})}\BibitemShut
  {NoStop}%
\bibitem [{\citenamefont {Colomer{-}de{-}Sim{\'o}n}\ \emph
  {et~al.}(2013)\citenamefont {Colomer{-}de{-}Sim{\'o}n}, \citenamefont
  {Serrano}, \citenamefont {Beir{\'o}}, \citenamefont {Alvarez-Hamelin},\ and\
  \citenamefont {Bogu{\~n}{\'a}}}]{colomer2013deciphering}%
  \BibitemOpen
  \bibfield  {author} {\bibinfo {author} {\bibfnamefont {P.}~\bibnamefont
  {Colomer{-}de{-}Sim{\'o}n}}, \bibinfo {author} {\bibfnamefont {M.~{\'A}.}\
  \bibnamefont {Serrano}}, \bibinfo {author} {\bibfnamefont {M.~G.}\
  \bibnamefont {Beir{\'o}}}, \bibinfo {author} {\bibfnamefont {J.~I.}\
  \bibnamefont {Alvarez-Hamelin}}, \ and\ \bibinfo {author} {\bibfnamefont
  {M.}~\bibnamefont {Bogu{\~n}{\'a}}},\ }\bibfield  {title} {\enquote {\bibinfo
  {title} {Deciphering the global organization of clustering in real complex
  networks},}\ }\href {http://dx.doi.org/10.1038/srep02517} {\bibfield
  {journal} {\bibinfo  {journal} {Sci. Rep.}\ }\textbf {\bibinfo {volume}
  {3}},\ \bibinfo {pages} {2517} (\bibinfo {year} {2013})}\BibitemShut
  {NoStop}%
\end{thebibliography}%
\end{document}